\documentclass[prb,twocolumn,reprint,amsmath,amssymb]{revtex4}
\usepackage{color}
\usepackage{bm}
\usepackage{dcolumn}
\begin{document}
\author{Kevin Leung}
\affiliation{Sandia National Laboratories, MS 1415,
Albuquerque, NM 87185, USA, {\tt E-mail: kleung@sandia.gov} 
}
\author{Rosy and Malachi Noked}
\affiliation{Department of Chemistry, Bar-Ilan University, Ramat Gan, 52900,
Israel, {\tt E-mail: rrosysharma@gmail.com},
{\tt E-mail: malachinoked@biu.ac.il} }
\date{\today}
\title{Anodic Decomposition of Surface Films on High Voltage Spinel Surfaces
-- Density Function Theory and Experimental Study}

\input epsf.sty
%\ssp
 
\begin{abstract}

Oxidative decomposition of organic-solvent-based liquid electrolytes
at cathode material interfaces has been identified as a main reason for
rapid capacity fade in high-voltage lithium ion batteries.  The evolution
of ``cathode electrolyte interphase'' (CEI) films, partly or completely
consisting of electrolyte decomposition products, has also recently
been demonstrated to be correlated with battery cycling behavior at high
potentials.  Using Density Functional Theory (DFT) calculations, the hybrid
PBE0 functional, and the (001) surfaces of spinel oxides as models, we examine
these two interrelated processes.  Consistent with previous calculations,
ethylene carbonate (EC) solvent molecules are predicted to be readily oxidized
on the Li$_x$Mn$_{2}$O$_4$ (001) surface at modest operational voltages,
forming adsorbed organic fragments.  Further oxidative decompostion of such
CEI fragments to release CO$_2$ gas is however predicted to require higher
voltages consistent with Li$_x$Ni$_{0.5}$Mn$_{1.5}$O$_4$ (LNMO)
at smaller $x$ values.  We argue that multi-step reactions, involving first
formation of CEI films and then further oxidization of CEI at higher
potentials, are most relevant to capacity fade.  
Mechanisms associated with dissolution or oxidation
of native Li$_2$CO$_3$ films, which is removed before the electrolyte
is in contact with oxide surfaces, are also explored.  

\end{abstract}

\maketitle

\section{Introduction}

The use of high voltage cathode materials like LiMn$_{1.5}$Ni$_{0.5}$O$_4$
spinel (LNMO) can contribute to significant increase in energy densities in
lithium ion batteries.\cite{ram_spinel,jow,gychen2014,manthiram15} 
Energy stored in batteries scale as $\Delta V^2$, where $\Delta V$ is
the voltage difference between anode and cathode. LNMO can operate at
$\sim$4.7-5.0~V vs.~Li$^+$/Li(s).  High nickel content layered lithium
nickel/manganese/cobalt (NMC) oxides also have the potential to expand
the voltage window.

\begin{figure}
\centerline{\hbox{(a) \epsfxsize=3.10in \epsfbox{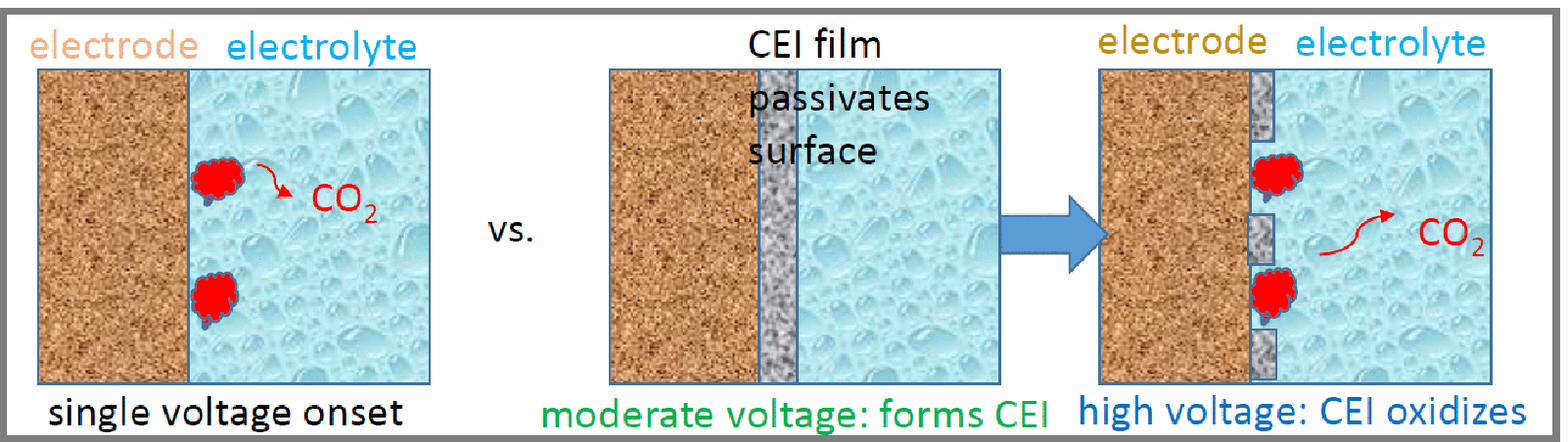} }}
\centerline{\hbox{(b) \epsfxsize=2.00in \epsfbox{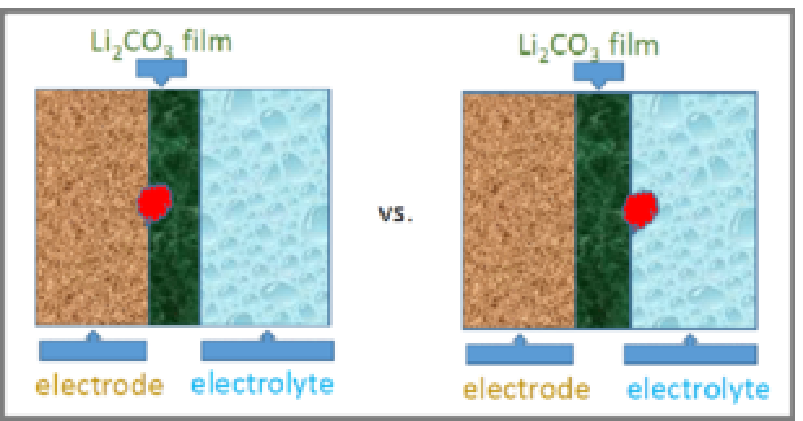} }}
\caption[]
{\label{fig1} \noindent
Two main science questions addressed in this work.
(a) One-step oxidative decomposition of organic-solvent-based electrolyte
at high voltage vs.~two- or multi-step degradation hypothesis, with the middle
panel at intermediate voltages relevant to $<$4.3~V conditions, and the right
panel at high voltages relevant to LNMO.  (b) Oxidation of Li$_2$CO$_3$ at
its interfaces with LNMO vs.~its reaction at the liquid electrolyte interface.
}
\end{figure}

One obstacle facing the deployment of high voltage cathode materials is
the apparent anodic instability of organic solvent molecules found in standard 
electrolytes (Fig~\ref{fig1}),
such as ethylene carbonate (EC) and dimethyl carbonate (DMC).\cite{xu} EC and
DMC oxidation has been proposed to lead to formation of thin cathode electrolyte
interphase (CEI) films.\cite{jarry,borodin_nature,shkrob2017,kanno1,kanno2,ram_spinel,oh,cei_pt,song,edstrom,aurbach01,aurbach99,novak}
While the existence of CEI films on cathode surfaces has long been confirmed
at modest voltages,\cite{edstrom,aurbach01,aurbach99,novak} questions
remain about the origin of CEI species.\cite{crosstalk,crosstalk1}
Although extensive spectroscopic and imaging studies have been conducted, the
structure and function of CEI on cathode oxide material surfaces remain poorly
understood.  To complicate matters, transition metal ion dissolution from
cathode oxide surfaces\cite{ram06} and surface phase transformation of layered
cathode oxides to form rock salts\cite{doeff,interfacetobulk} can occur at
the same time.  Protective coating has been applied with some
success.\cite{ald0,ald1,ald2,ald3,ald4,ald5} Electrolytes tailored for
high voltages can also circumvent this
problem,\cite{borodin_nature,shkrob2017,sulfone} often at the cost of
higher electrolyte viscosity and material expense.

There is increasing experimental evidence that CEI components on cathode
oxides are not static; they continue to evolve and/or become further oxidized
as cycling proceeds to high voltages.\cite{novak,oh,matsui,shendillon1,shendillon3,amatucci1,sargita,ma}  This finding is not limited to the cathode side;
evolution of SEI physical properties and chemical composition on anode
surfaces has also been reported.\cite{anode_sei1,anode_sei,batt,perla}
Recent differential  and online electrochemical mass spectroscopy (DEMS, OEMS)
measurements have proven extremely useful for correlating gas release with
voltage changes.\cite{novak2016,gasteiger2018,gasteiger2017,gasteiger2015,gasteiger2014,abruna,janek2016,behm}  
The onset of CO$_2$ release from LNMO cells has been reported to be $\sim$4.6
to~5.5~V, except for open circuit and first-cycle contributions.  CO$_2$
release from layered NMC occurs at lower potentials and has been linked to
reactive oxygen release.\cite{gasteiger2018}

Regarding theoretical modeling, anion-mediated EC oxidation has been predicted
to occur in bulk solutions at about 5~V potentials.\cite{borodin1,borodin17} 
EC is also predicted to undergo a one-step CO$_2$ release reaction on
LNMO (001) surfaces.\cite{tateyama} However, there is substantial
computational evidence that EC and DMC molecules already react with
cathode oxide surfaces at more modest 
voltages,\cite{leung1,leung2,borodin2015,giordano1,giordano2,musgrave1,tebbe,morgan,bal19,li17,evenstein,intan19,benedek,evenstein}
especially on the surfaces of nickel-free manganese spinels (LiMn$_{2}$O$_4$,
or LMO) which do not normally operate beyond $\sim$4.3~V.  Strictly speaking,
the predicted reactions are interfacial.  Long
range electron transfer into the liquid electrolyte is not yet predicted;
instead direct contact between electrolyte and oxide surfaces is
required.\cite{note1}   These predictions are inconsistent with
the view that high voltages are needed to oxidize the electrolyte.  

Based on these computational and experimental findings, we adopt
an alternative view of high voltage cathode interfacial reactions
which has already been suggested for oxidation on Pt surfaces\cite{cei_pt}
(Fig.~\ref{fig1}a).  (1) Solvent molecules in standard organic battery
electrolyte first react with cathode oxide surfaces at modest voltages during
charging.  This occurs even on non-high-voltage materials like
Li$_x$Mn$_2$O$_4$. The oxidation generates adsorbed fragments which, to some
extent, ``passivates'' surface reaction sites.  (2) At higher voltages during
charging, these CEI products become oxidized and are removed, releasing CO$_2$
gas and leading to uncontrolled further electrolyte decomposition.  We argue
that the latter event is more responsible for rapid capacity fade
(Fig.~\ref{fig1}a).  

In this work, solvent decomposition products previously predicted to adsorb
on LMO\cite{leung1,leung2} constitute our ``CEI'' model components.  Until
these species desorb, dissolves, and/or are themselves oxidized, they should
sterically passivate cathode surfaces.  We focus on
these organic fragments, and omit inorganic CEI components like LiF and
Li$_w$P$_x$O$_y$F$_z$, related to PF$_6^-$ decomposition.  Such inorganic
species are present\cite{edstrom,aurbach01,aurbach99,zhang2019,zhao2019} but
neither release CO$_2$ gas nor have been predicted to be oxidized at any
reasonable voltages.  Indeed, PF$_6^-$ decomposition products have been
suggested to partially passivate LNMO.\cite{novak2016}

We apply Density Functional Theory calculations to illustrate the hypothesis
(Fig.~\ref{fig1}a).  We adopt the (001) surfaces of LMO and high voltage spinel
LNMO as model systems, and compute the thermodynamics and kinetics associated
with degradation reactions.  First we re-examine the initial EC oxidation
steps\cite{leung1} using the hybrid PBE0 functional,\cite{pbe0} which is
generally more accurate for reaction barriers\cite{pbe_barrier,wtyang} than
the DFT+U method\cite{dftu1} widely used in the literature,\cite{leung1,leung2,borodin2015,giordano1,giordano2,musgrave1,tebbe,morgan,bal19,li17,evenstein,intan19,benedek}
but is far more computationally costly.  Next we consider the further oxidation
of these initial products on LNMO, and show that lower Li and higher Ni
contents, which correspond to higher voltages, are needed to lower the reaction
barriers sufficiently to activate further reactions that release CO$_2$.  Due
to the computational cost, PBE0 is only used to examine the key reaction steps.
One key finding is that DFT+U can significantly underestimate reaction barriers 
compared to PBE0, yielding qualitative changes in mechanistic interpretations.
Online electrochemical mass spectroscopy (OEMS) measurements are also conducted
to support the theoretical results.

Other cathode interfacial evolution phenomena are related to the above
discussions.   First, the native Li$_2$CO$_3$ film covering most
as-synthesized cathode oxide materials must somehow disappear or be ruptured
before the elcetrolyte can react with them.  It is widely accepted
that native Li$_2$CO$_3$ films readily form on cathode oxide surfaces upon
exposure to CO$_2$ in air after synthesis.\cite{lbl}  Li$_2$CO$_3$ is known
to be oxidized to yield CO$_2$ at $\sim$4.2~V\cite{lbl,luntz} or
above.\cite{gasteiger2013}  The carbonate content on CEI films on layered
NMC has indeed been reported to fluctuate as cycling
continues.\cite{matsui,shendillon1,shendillon3,amatucci1,luntz}
Some studies suggest Li$_2$CO$_3$ dissolves due to reactions with LiPF$_6$
degradation products like HF.\cite{amatucci1}  However, Li-air battery studies,
which do not involve PF$_6^-$ that generates HF, also report Li$_2$CO$_3$
oxidation.\cite{luntz,gasteiger2013}  Using the DFT+U method and storage
condition experiments, we perform exploratory studies to investigate
whether oxidative reactions at the liquid electrolyte/Li$_2$CO$_3$ interface
occur more readily
than at the Li$_2$CO$_3$/LNMO (001) interface (Fig.~\ref{fig1}b).  Another
interesting phenomenon is one version of ``cross-talk''  where SEI fragments
diffuse from the anode to the cathode\cite{crosstalk,crosstalk1} and become
oxidized.\cite{sargita}  Cross-talk products have variable
structures,\cite{crosstalk1} and they will not be considered herein.
However, the principles and methods used in this work can also apply to
study this phenomenon.

\section{Methods}
\label{method}

We estimate mean reaction rates using the standard transition state theory
rate equation
\begin{equation}
1/t_{\rm ave}  = k_o \exp (-\Delta E^*/k_{\rm B}T), \label{eq1}
\end{equation} 
where $\Delta E^*$ is the activation energy, $k_o$=10$^{12}$/s is a standard
kinetic prefactor, and $k_{\rm B}T$ is the thermal energy at room temperature.
Any step in the proposed reaction mechanisms is considered favorable if
it is exothermic ($\Delta E$$<$0) and if $t_{\rm ave}$ is less than one hour
-- which translates into $\Delta E^*$$<$$\sim$1~eV.\cite{batt} Eq.~\ref{eq1}
assumes T=0~K and ignores entropy which is small in most cases.  As will be 
discussed, in a few relevant cases gas phase translational or vibrational
entropic effects are added post-processing.

$\Delta E$ and $\Delta E^*$ are computed using T=0~K static ultrahigh vacuum
(UHV) condition DFT calculations, conducted using periodically replicated
simulation cells and the Vienna Atomic Simulation Package (VASP) version
5.3.\cite{vasp1,vasp1a,vasp2,vasp3}  A 400~eV planewave
energy cutoff and a 10$^{-4}$~eV convergence criterion are enforced.  
Antiferromagnetic ordering is imposed on Li$_x$Ni$_{0.5}$Mn$_{1.5}$O$_4$
and Li$_x$Mn$_{2}$O$_4$.  

Most DFT calculations herein apply the Perdew-Burke-Ernzerhof (PBE)
functional\cite{pbe} with the Hubbard (DFT+U) augmentation.\cite{dftu1}
The $U$ and $J$ values associated with DFT+U depend on the orbital
projection scheme and DFT+U implementation details; here $(U-J)=$4.84~eV,
5.96, and 3.30~eV for Mn, Ni, and Co in accordance with the
iterature.\cite{zhou,persson14,liverpool1,islam}  Co-based materials are
discussed only in the S.I.  The PBE functional that underlies DFT+U
calculations tends to underestimate reaction barriers\cite{pbe_barrier}
because of localization errors.\cite{wtyang}  The hybrid PBE0 functional,
generally considered more accurate than PBE for barriers,\cite{pbe_barrier} is
applied to key reaction steps as spot checks.\cite{pbe0}  Although fewer
in number, these PBE0 calculations constitute the main results in this work.
See the S.I.~for rationale for choosing PBE0 over other hybrid functionals.
In slab geometry DFT+U calculations, the standard dipole correction is
applied.\cite{dipole} It is found that this correction is on the order
of meV, and it is omitted in PBE0 calculations.  

Reaction barriers ($\Delta E^*$) are computed using
the climbing image nudged elastic band (NEB) method.\cite{neb} 
PBE0-based NEB calculations require far more ionic steps than DFT+U NEB.  The
reason is that, in the relevant oxidative reactions, protons and $e^-$ are
transferred simultaneously.  The (spatial) radius of convergence associated
with PBE0 NEB is small because the $e^-$ being transfered is more localized
than in DFT+U predictions.  Therefore good initial guesses are needed to
converge PBE0 NEB.  Using DFT+U NEB configurations as guesses routinely
fails to yield PBE0 barriers.  

2$\times$2$\times$2 and 2$\times$2$\times$1 Brillouin zone sampling schemes
are adopted for LMO and LNMO bulk and surface unit cells.  The DFT+U lattice
constants for LMO and LNMO are predicted to be 8.40~\AA\, and 8.30~\AA,
respectively.  PBE0 simulation cells are assumed to have the same cell
dimensions.  1$\times$1 surface cells are created by exposing (001)
surfaces.\cite{lmo111,persson14,persson13,vitaly}  The resulting slabs have
Li$_n$Ni$_4$Mn$_{16}$O$_{40}$ or Li$_n$Mn$_{20}$O$_{40}$ stoichiometries and
no net dipole moment normal to the surface.  The slight reduction in the 
Ni/Mn ratio in our LNMO surface models, compared to the canonical 
Li$_x$Ni$_{0.5}$Mn$_{1.5}$O$_{4}$, is the result of using a fairly thin slab.

For the 1$\times$1 (001) surface cells used, PBE0 is up to 100 times more
costly than DFT+U for each ion step.  Therein lies the advantage of using
a small surface cell: the PBE0 functional can be applied more readily.
Furthermore, the energies of all configurations with different Ni positions can
be enumerated (at least when using the DFT+U method).  With $x\approx 1$
(i.e., under synthetic conditions), it is most energetically favorable to
have one Ni on each surface and two Ni in the interior.  To check system
size effects, some DFT+U calculations associated with the oxidation of EC
fragments apply 1$\times$2 surface supercells along
with 2$\times$1$\times$1 Brillouin sampling.  These results, and justification
for using a single adsorbed molecule at T=0~K to model anodic decomposition,
are discussed in the S.I.

DFT+U-based vibrational frequency calculations are conducted to confirm that
one of the configurations predicted to be a transition state (TST) indeed has
only one unstable mode.  The frozen phonon method is applied, with only the
EC fragment allowed to move.  In other words, oxide ions are assumed to be
infinitely heavy.  The vibrational contribution to the free energy difference
between initial and transition states is estimated using a harmonic
approximation,
\begin{eqnarray}
\Delta \Delta A &=& \Delta A_{\rm TST}-\Delta A_{\rm initial} ; \nonumber \\
\Delta A_y &=& \sum_{i=1}^{n_y} \hbar \omega_i + k_{\rm B}T 
		\log [1-\exp(-\hbar \omega_i /k_{\rm B}T ) ] \label{eq2}
\end{eqnarray}
where $y$=initial or TST, $n_y$=30 or~29 in the two cases respectively,
$\omega_i$ is the $i$th vibrational frequency in the EC fragment, and $\hbar$
is Planck's constant.

Calculations on Li$_2$CO$_3$/LNMO interfaces apply 1$\times$3 LNMO surface
supercells along with 2$\times$1$\times$1 Brillouin sampling.  More details
are provided in the S.I.  Reactions between Li$_2$CO$_3$ films and EC molecules
apply 2$\times$2$\times$1 Brilloin zone zampling,
8.34$\times$10.02$\times$28~\AA$^3$ simulation cells with 16 Li$_2$CO$_3$
units, and 4~Li removed from the surface.

The net electronic spin on each transition metal ion is examined to determine
their charge states.  Mn(II), Mn(III), and Mn(IV) are identified as Mn ions
which exhibit net spins of $\sim$4.6, $\sim$4.0, and $\sim$3.3 (all to within
$\pm$0.3 unit), as reported by the VASP code using its default PAW orbital
settings.  For added verification, the maximally localized Wannier orbitial
method\cite{wannier} is applied to locate the center of all occupied
orbitals within 0.3~\AA\, of each transition metal ion.  By counting the 
number of occupied $d$-Wannier orbitals centered around each Mn/Ni,
the charge state can be unambiguously assigned.

Regarding experiments: for making the composite electrodes, a uniform slurry
was prepared by mixing 84\% LNMO, 6\% CMC, and 10\% C65. The contents were
thoroughly mixed using Thinky Mixer (ARV-310/ARV-310LED) at 2000~rpm and
40~kPa for 3~minutes.  The prepared slurry was then coated on a clean and
polished Al foil using a doctor's blade adjusted for 60~$\mu$m thickness.
The coated sheet of aluminum foil was then heated at 100$^o$C followed by
rolling to ensure complete removal of trapped air or solvent. Vacuum dried
electrodes with 12~mm diameter were used for the OEMS study.

	For online electrochemical mass spectrometry, the OEMS cells were
assembled using Li (14~mm diameter) as anode and LNMO composite electrode
(12~mm diameter) as cathode. 2 poly-propylene separators (29~mm diameter)
were used between the two
electrodes with 100 $\mu$L of LP30 (1M LiPF$_6$ in EC:DMC (1:1)) electrolyte.
The cell was connected to OEMS (HPR-40, Hiden analytical) using a
micro-capillary inlet with a sample rate of 12 $\mu$L/min. The galvanostatic
charge/discharge was carried out using VSP- potentiostat (Bio-logic
Science instruments) in a potential window of 3.5~V to 5.0~V at a rate
of C/10. For the in-operando measurements of the evolved gases as a
function of applied potential in the real-time frame, Mid mode was selected
for H$_2$ (m/z = 2), CO$_2$ (m/z = 44), and CO and C$_2$H$_4$ (m/z = 28) gases.

\section{Results and Discussions}

\subsection{Revisiting EC oxidation on LMO using DFT/PBE0}
\label{ec}

\begin{figure}
%\centerline{\hbox{ (a) \epsfxsize=1.50in \epsfbox{fig2a.ps}  
%                   (b) \epsfxsize=1.50in \epsfbox{fig2b.ps}  }}
%\centerline{\hbox{ (c) \epsfxsize=1.50in \epsfbox{fig2c.ps}  
%                   (d) \epsfxsize=1.50in \epsfbox{fig2d.ps}  }}
\centerline{\hbox{ (e) \epsfxsize=3.00in \epsfbox{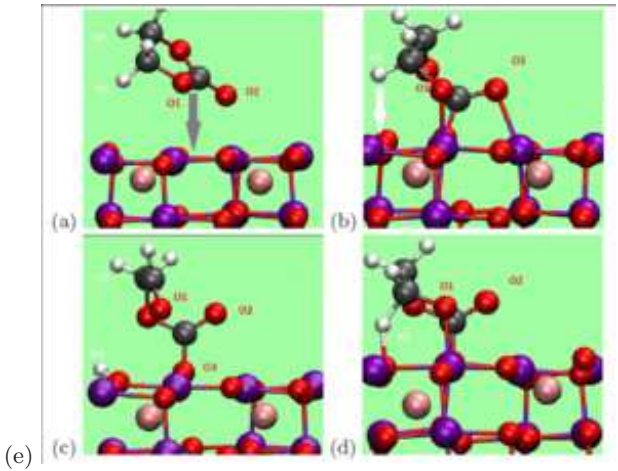} }}
\caption[]
{\label{fig2} \noindent
PBE0-predicted configurations associated with EC decomposition on
Li$_{0.6}$Mn$_{2}$O$_4$ (001).  (a) Intact EC; (b) partially decomposed
but unoxidized EC fragment on Li$_{0.6}$Mn$_{2}$O$_4$ (001); (c)
oxidized EC fragment, $\Delta E$=-1.91 eV (PBE0) relative to (a);
(d) transition state configuration ($\Delta E^*$=+1.05~eV, PBE0).
%(e) derived from (d) with an O$^{2-}$ anion pulled out of the surface.
Purple, pink, red, grey, and white spheres represent Mn, Li, O, C, and H atoms.
}
\end{figure}

Our previous computational work,\cite{leung1,leung2} using the DFT+U method,
has shown that ethylene carbonate (EC) decomposition occurs readily on LMO
(001) and (111). In particular, 40\% charged LMO (Li$_{0.6}$Mn$_2$O$_4$)
already oxidizes EC molecules on its (001) surface.  The rate-determining,
oxidative step releases $\Delta E$=-2.0~eV.  The reaction barrier is
predicted to be only $\Delta E^*$=0.56~eV.  According to Eq.~\ref{eq1},
a reaction with this small $\Delta E^*$ occurs in
millisecond time scales.  Thus Li$_{0.6}$Mn$_2$O$_4$, which should operate
below 4.3~V, is predicted to rapidly oxidize EC molecules.  Higher voltages
are not needed.  The more accurate PBE0 mehthod has also been used to 
calculate $\Delta E$.\cite{leung1} However, PBE0 reaction barriers
($\Delta E^*$) have not been previously reported.

Fig.~\ref{fig2} describes new PBE0 $\Delta E^*$ predictions computed
using the climbing image nudged elastic band method (NEB).\cite{neb} First
Fig.~\ref{fig2}a depicts an intact EC on LMO (001) surface.  Fig.~\ref{fig2}b-d
depict the initial, final, and barrier configurations associated with the
rate-determining oxidative step in the initial decomposition of EC molecules.
Two $e^-$ and a H$^+$ are transferred.  $\Delta E$ is $-1.75$~eV, which is only
0.25~eV less exothermic than the DFT+U value.\cite{leung1}  $\Delta E^*$ at
the barrier top (Fig.~\ref{fig2}d) is found to be 1.05~eV higher in energy than
Fig.~\ref{fig1}b.  This PBE0 $\Delta E^*$ is 0.49~eV larger than DFT+U
value, in accordance with our expectation that PBE0 barriers are generally
higher.\cite{pbe_barrier}  This PBE0 $\Delta E^*$ value should be
more accurate.\cite{pbe_barrier}  1.05~eV is still consistent with a
fast degradation reaction at room temperature, especially because the large
zero point energy (ZPE) correction associated with proton motion has so far
been neglected.  ZPE contribution at T=300~K can be estimated using
DFT+U configurations, DFT+U frozen phonon calculations which yield a single
unstable vibrational mode at the transition state, and Eq.~\ref{eq2}.
We find that ZPE lowers $\Delta E^*$ (or more appropriately $\Delta G^*$,
the free energy barrier) by~0.16~eV, to only 0.89~eV.  This value is consistent
with 4~reactions per hour (Eq.~\ref{eq1}), well within battery time scales.
We conclude that CEI products readily form on moderate voltage LMO (001).

Given the cost of PBE0 barrier calculations, we have not performed the same
two $e^-$, one H$^+$ transfer calculation on LNMO (001).  Since LNMO operates
at higher voltages, and should yield more exothermic, faster reactions than
corresponding ones on LMO, it is reasonable to assume the reactions of
Fig.~\ref{fig2} also readily occur on LNMO (001) at the same Li-content.

\subsection{Further Oxidation on LMO and LNMO: DFT+U Predictions}

\begin{figure}
%\centerline{\hbox{ (a) \epsfxsize=1.50in \epsfbox{fig3a.ps}  
%                   (b) \epsfxsize=1.50in \epsfbox{fig3b.ps}  }}
%\centerline{\hbox{ (c) \epsfxsize=1.50in \epsfbox{fig3c.ps}  
%                   (d) \epsfxsize=1.50in \epsfbox{fig3d.ps}  }}
%\centerline{\hbox{ (e) \epsfxsize=1.50in \epsfbox{fig3e.ps}
%                   (f) \epsfxsize=1.50in \epsfbox{fig3f.ps}  }}
%\centerline{\hbox{ (g) \epsfxsize=1.50in \epsfbox{fig3g.ps}
%                   (h) \epsfxsize=1.50in \epsfbox{fig3h.ps}  }}
\centerline{\hbox{ \epsfxsize=3.00in \epsfbox{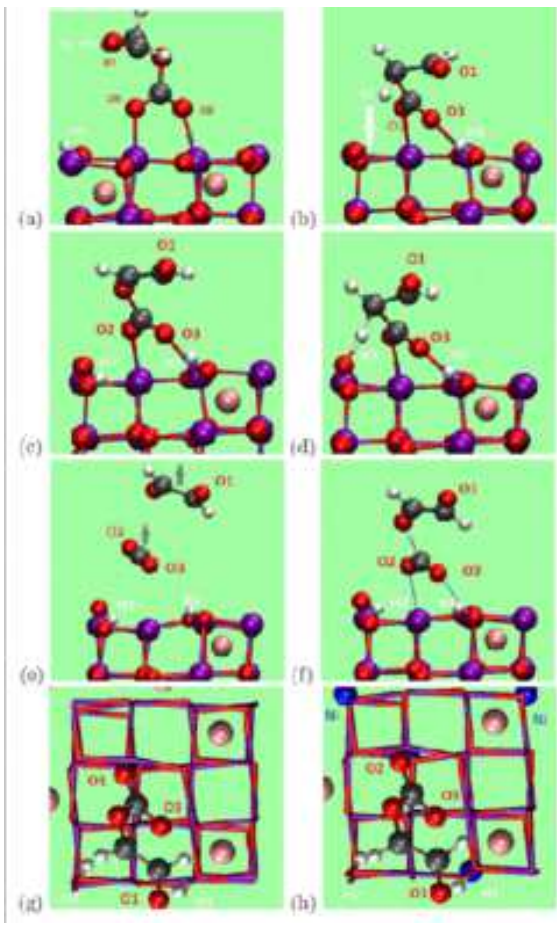} }}
\caption[]
{\label{fig3} \noindent
LMO (001) configurations obtained using DFT+U calculations, but PBE0
configurations look superficially similar.
(a) Surface O$^{2-}$ anion coordinated to C moves out of the surface
compared with Fig.~\ref{fig2}c;
(b) rotating panel (a) configuration and removing a Li;
(c) transferring a second H to surface;
(d) H transfer transition state; 
(e) breaking C-O bond; (f) C-O cleavage transition state;
(g) top view of panel (c); 
(h) same as (g) but for LNMO, with Ni depicted as blue spheres.  
Purple, pink, red, grey, and white spheres represent Mn, Li, O, C, and H atoms.
}
\end{figure}

As discussed in Ref.~\onlinecite{leung1}, efforts to break other bonds in this
partially oxidized EC fragment (Fig.~\ref{fig2}c) have yielded endothermic
reactions.  Hence this fragment has been assumed to persist during battery
charging.  However, a hitherto unexamined pathway proves favorable
when using the DFT+U method.  We stress, particularly to experimentalist
readers, that more accurate PBE0 calculations presented in the next section
suggest that DFT+U overestimates reactivities.  Nevertheless, it is useful
to first report DFT+U predictions.

The Fig.~\ref{fig3}a configuration spontaneously emerges from Fig.~\ref{fig2}c
in finite temperature DFT+U-based molecular dynamics simulations,\cite{leung1}
and involves lifting one O$^{2-}$ anion out of the LMO surface.
Static DFT+U calculations predict that it is 0.10~eV more favorable than
Fig.~\ref{fig2}c.  We rotate this organic fragment so one of its H-atoms 
faces the oxide surface, remove one more Li to partially offset the implicit
voltage decrease due to transfer of $e^-$ from EC to the oxide, and use
the resulting configuration (Fig.~\ref{fig3}b) as the re-starting point.

The next favorable step involves the transfer of a second H$^+$ and an $e^-$
to the LMO surface (Fig.~\ref{fig3}c).  It is exothermic by $\Delta E$=-0.46~eV
using the DFT+U method  (Fig.~\ref{fig4}a). 
The DFT+U transition state (Fig.~\ref{fig3}d) exhibits $\Delta E^*$=+0.60~eV
relative to Fig.~\ref{fig3}b; this value is similar to that in the proton
transfer step (Fig.~\ref{fig2}d) when using DFT+U.\cite{leung1}

Once this second hydrogen is transferred, the cleavage of
the bond between the carbonyl carbon and an oxygen originally in the
EC 5-member ring becomes favorable by $\Delta E$=-0.15~eV relative to
Fig.~\ref{fig3}c according to DFT+U.  Breaking this bond leads to the
release of both a CO$_2$ and a glyoxal (C$_2$H$_2$O$_2$) molecule
(Fig.~\ref{fig3}e).  Without the prior H-atom transfer step,
CO$_2$ elimination would have been accompanied by the release of a
CHOCH$_2$O radical, which is less energetically favorable, although a similar
reaction appears favorable on LNMO,\cite{tateyama} not LMO.  It
results in a 4-coordinated Mn(II) ion in Fig.~\ref{fig3}e.  The DFT+U-predicted
barrier configuration (Fig.~\ref{fig3}f) exhibits $\Delta E^*$=0.46~eV relative
to Fig.~\ref{fig3}c, and has a 1.95~\AA\, C-O bond length.  C$_2$H$_2$O$_2$
is known to be a polymerizing agent.  It is likely to react either with the
oxide surface or other EC molecules in the solvent, and is not
expected to be detected after cycling.  

%\begin{table}
%\begin{tabular}{||l|l|r|r|r||r|r|r||} \hline
%surface & method & Fig.~\ref{fig2}b & Fig.~\ref{fig3}c & Fig.~\ref{fig3}d &
%	Fig.~\ref{fig3}c & Fig.~\ref{fig3}e & Fig.~\ref{fig3}f \\ \hline 
%LMO(001) & DFT+U & 0.00 & -0.46 & +0.60 & 0.00 & -0.15 & +0.46 \\
%LNMO(001) & DFT+U & 0.00 & -0.41 & +0.58 & 0.00 & -0.32 & +0.32 \\
%LMO(001) & PBE0 & 0.00 & -0.62  & NA & 0.00 & +0.62 & NA \\
%LNMO(001) & PBE0 & 0.00 & -0.43 & NA & 0.00 & +0.08 & +1.22 \\
%%LNMO(001)$^\dagger$ & DFT+U & 0.00 & -0.72 & NA & 0.00 & -0.82 & +0.77 \\
%LNMO(001)$^\dagger$ & PBE0 & 0.00 & -0.72 & NA & 0.00 & -0.82 & +0.77 \\
%\hline
%\end{tabular}
%\caption[]
%{\label{table1} \noindent
%Relative energies and energy barriers, in eV, associated with further reactions
%of EC fragments on LMO and LNMO (001) surfaces.  The EC configurations
%on LNMO (001) are like those for LMO (Fig.~\ref{fig3}).  The Fig.~\ref{fig3}d
%intermediate state is a product of the first reaction and is the reactant in
%the next; it listed twice with its $\Delta E$ reset to 0~eV in the second
%step.  All systems have 50\% Li content except the one marked by
%a dagger, which has 20\% Li content.
%}
%\end{table}

DFT+U predicted EC fragment configurations on LNMO (001) are similar to
those on LMO (001), and only one configuration is depicted (Fig.~\ref{fig3}h).
The only qualitative difference with LMO is that the 4-coordinated Mn in the
final configuration on LNMO (Fig.~\ref{fig3}e for LMO) remains a Mn(III); a
nickel ion is reduced instead of a Mn(III) following oxidation of the organic
fragment.  The final step $\Delta E^*$=+0.32~eV is lower than the corresponding
LMO value (Fig.~\ref{fig4}).  Because the surface Ni introduces
heterogeneity with respect to the EC binding site, we have translated the
organic fragment in Fig.~\ref{fig3}h about the diagonal in the surface cell
containing the surface transition metal ions, and find that the configuration
depicted is within 0.03~eV of the most stable among 8 choices we have
considered.

The main problem with these DFT+U results is that the predicted $\Delta E$
and $\Delta E^*$ for reactions on LNMO surfaces are similar to those on
LMO (Fig.~\ref{fig4}a).  Both oxidation steps subsequent to the LNMO equivalent
of Fig.~\ref{fig3}c are exothermic, and the average reaction times are much
less than one hour.  With all adsorbed organic fragments removed from
their surfaces, LMO and LNMO should continuously react with solvent
molecules and evolve CO$_2$ at LMO operating potentials ($<4.3$~V).  
Experimentally, CO$_2$ release is observed with LNMO at high voltage, but
{\it not} with LMO at lower voltages after the first cycle.\cite{novak2016,gasteiger2018,gasteiger2017,gasteiger2015,gasteiger2014,abruna,janek2016,behm}
DFT+U fails to distinguish LMO spinel and high-voltage LNMO spinel.  Doubling
the size the the simulation cell and adding more EC molecular fragments does
not resolve this discrepancy between modeling and experiments (S.I.).
The absence of molecular fragments on the cathode oxide at intermdediate
voltages also seems inconsistent with two apparent CO$_2$ onset
potentails\cite{novak2016}  (see Sec.~\ref{oems} below).  

\subsection{Further Oxidation on LMO and LNMO: PBE0 Predictions}

\begin{figure}
\centerline{\hbox{ \epsfxsize=3.00in \epsfbox{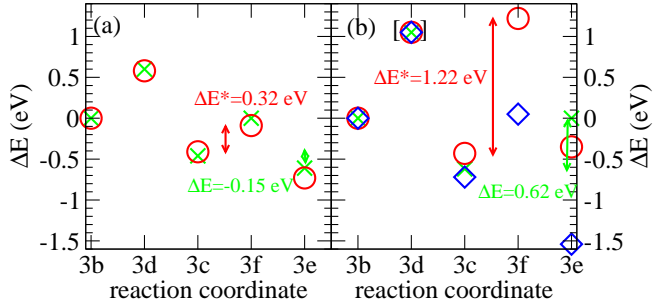} }}
\caption[]
{\label{fig4} \noindent
Relative energies and energy barriers, in eV, associated with further reactions
of EC fragments on LMO and LNMO (001) surfaces.  Panels (a) and (b) refer to
DFT+U and PBE0 calculations, respectively.  Green, red, and blue (only in (b))
are for Li$_{0.5}$Mn$_2$O$_4$, Li$_{0.5}$Ni$_{0.5}$Mn$_{1.5}$O$_4$, and
Li$_{0.2}$Ni$_{0.5}$Mn$_{1.5}$O$_4$ respectively.  The $x$-axis refers to
Fig.~\ref{fig3} panels; LNMO configurations are similar to LMO ones in
Fig.~\ref{fig3}.  The bracket in (b) means that the PBE0 proton transfer
$\Delta E^*$ are not calculated, but are assumed to be similar to that
associated with Fig.~\ref{fig2}d.  The two-sided arrows indicate the two
key differences between (a) DFT+U and (b) PBE0.  PBE0 predicts that the
final reaction on Li$_{0.5}$Ni$_{0.5}$Mn$_{1.5}$O$_4$ has a much higher
$\Delta E^*$ than DFT+U, and that $\Delta E$ for the final reaction
on Li$_{0.5}$Mn$_2$O$_4$ is much more unfavorable than DFT+U.
}
\end{figure}

The PBE0 method is next applied to key reaction steps to distinguish LNMO
from LMO.  The PBE0-predicted configurations are qualitatively similar to
those in Fig.~\ref{fig3}, and are depicted only in a few cases.  But the
energetics are substantially different (Fig.~\ref{fig4}b).

\begin{figure}
%\centerline{\hbox{ (a) \epsfxsize=1.50in \epsfbox{fig5a.ps} 
%                   (b) \epsfxsize=1.50in \epsfbox{fig5b.ps}  }}
%\centerline{\hbox{ (c) \epsfxsize=1.50in \epsfbox{fig5c.ps} 
%                   (d) \epsfxsize=1.50in \epsfbox{fig5d.ps}  }}
%\centerline{\hbox{ (e) \epsfxsize=1.50in \epsfbox{fig5e.ps} 
%		   (f) \epsfxsize=1.50in \epsfbox{fig5f.ps}  }}
\centerline{\hbox{ \epsfxsize=3.00in \epsfbox{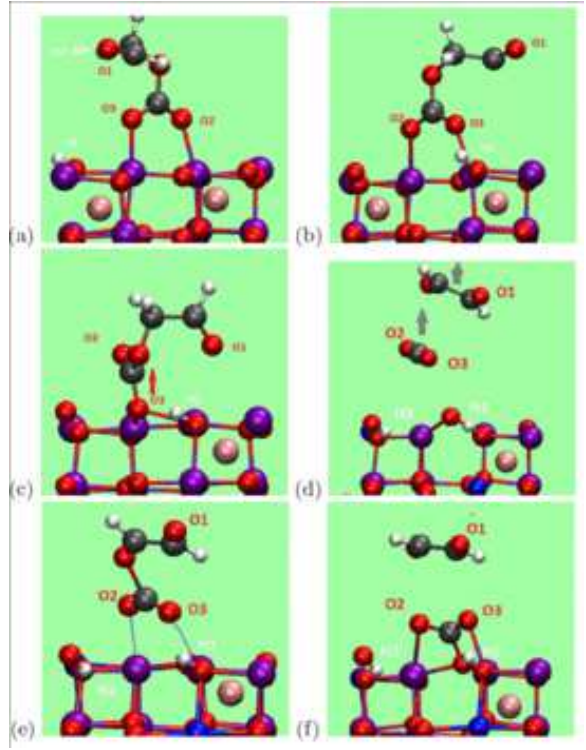} }}
\caption[]
{\label{fig5} \noindent
Selected LMO and LNMO configurations obtained using PBE0 calculations.
(a)-(b) Before/after dragging O$^{2-}$ anion off
Li$_{0.6}$Mn$_{2}$O$_4$ (001) surface.
(c) Removing one Li from panel (a).
(d)-(e) Final configuration and transition state for EC oxidation on
Li$_{0.5}$Ni$_{0.5}$Mn$_{1.5}$O$_4$, respectively.  
(f) CO$_2$ re-adsorbs on Li$_{0.5}$Ni$_{0.5}$Mn$_{1.5}$O$_4$.
The Ni cation on the surface in panels (d)-(f) is slightly obscured by
the leftmost OH group; see Fig.~\ref{fig3}h for clarity.
}
\end{figure}

First we revisit the transition between configurations Fig.~\ref{fig2}c
and Fig.~\ref{fig3}a, which does not involve electron transfer.  At this
Li$_{0.6}$Mn$_2$O$_4$ stoichiometry, DFT+U predicts $\Delta E$=-0.10~eV.
In contrast, the PBE0 functional predicts that $\Delta E$=+0.16~eV.  The PBE0
starting and ending configurations are depicted in Fig.~\ref{fig5}a-b.  It
is reasonable to assume that the PBE0 value is more accurate.  However, for
the purpose of modeling further oxidizing reactions, we should remove more
Li atoms to partially offset the $e^-$ transferred from EC to the LMO
slab so as to maintain high voltages.  Next we eliminate the Li directly
coordinated to the surface O$^{2-}$ anion being dragged by the EC fragment
off the surface (labeled ``O'' in Fig.~\ref{fig5}c) to yield a
Li$_{0.5}$Mn$_2$O$_4$ stoichiometry.  Now the equilivalent of Fig.~\ref{fig5}b,
but with one less Li, becomes exothermic ($\Delta E$=-0.25~eV, not show in
Fig.~\ref{fig5}) relative to Fig.~\ref{fig5}c, suggesting that this step
should proceed.  The overall Li-removal from Fig.~\ref{fig5}a yields a
+4.70~V equilibrium voltage if one takes into account lithium metal cohesive
energy.  (Here ``equilibrium'' means ``electrochemical
equilibrium'';\cite{leung3} see Fig.~\ref{fig7}b below.)  Thus a 4.70~V
potential appears necessary for further EC oxidation on LMO.  By comparison,
this requirement is found to be only 4.33~V in DFT+U calculations.

The next deprotonation step (Fig.~\ref{fig3}b $\rightarrow$ Fig.~\ref{fig3}c)
is predicted to be exothermic using both DFT+U and PBE0 functionals
(Fig.~\ref{fig4}a-b).  Since this is another proton transfer, we assume that
the PBE0 $\Delta E^*$ is not dissimilar to the $\Delta E^*$ computed in
Fig.~\ref{fig2}d for that H-transfer (1.05~eV, bracketed in Fig.~\ref{fig4}b),
and omit this step from PBE0 consideration.

Instead we focus on the final step.  For Li$_{0.5}$Mn$_2$O$_4$, PBE0
predicts that the Fig.~\ref{fig3}c $\rightarrow$ Fig.~\ref{fig3}e reaction,
associated with CO$_2$ and C$_2$H$_2$O$_2$ release, exhibits $\Delta E$=0.62~eV.
This is far more endothermic than the DFT+U prediction (Fig.~\ref{fig4}a).
We speculate that the DFT+U parameter for Mn is fitted to Mn(III)/Mn(IV)
solid state electrochemistry and may overestimate the stability of the Mn(II)
found in Fig.~\ref{fig3}.

Fig.~\ref{fig5}d depicts the product of the CO$_2$ release reaction on
Li$_{0.5}$Ni$_{0.5}$Mn$_{1.5}$O$_4$, instead of LMO, at the same Li content.
While the EC fragment configuration superficially resembles that on LMO
surfaces, the presence of Ni makes this reaction far less endothermic
($\Delta E$=0.08~eV Fig.~\ref{fig4}).  CO$_2$ release should be accompanied
by the typical favorable entropy of $\sim$0.4-0.5~eV at T=300~K at estimated
gas pressure of 0.1~atm.,\cite{ma} which offsets the 0.08~eV endothermicity.
In fact, all steps of further EC oxidation we have examined so far with the
PBE0 method (Fig.~\ref{fig3}, Fig.~\ref{fig5}) are more favorable on LNMO (001)
than LMO (001) slabs at the same 50\% Li content (Fig.~\ref{fig4}a).  Just by
considering energetic differences, the PBE0 functional (unlike DFT+U)
already distinguishes oxidation on LNMO from that on LMO.  

Next we examine the kinetics associated with CO$_2$ release on LNMO.
The $\Delta E^*$ predicted at the PBE0 transition state (Fig.~\ref{fig5}e)
is larger than the DFT+U value by a factor of 3.5 (Fig.~\ref{fig4}).  
Whereas the DFT+U $\Delta E^*$ is consistent with a fast reaction, the PBE0
value of $\Delta E^*$=1.22~eV suggests that the organic fragment shown in
Fig.~\ref{fig3}d should persist over beyond battery operation time scales on
Li$_{0.5}$Ni$_{0.5}$Mn$_{1.5}$O$_4$ at T=300~K. (Note that the Fig.~\ref{fig5}e
transition state does not involve proton transfer, and a large zero point
correction is not expected.) We have not repeated the similar reaction on LMO
surface because LMO is expected
to be even less oxidative than high-voltage LNMO at the same Li content.  It
is perhaps not surprising that this complex transition state proves
problematic for the DFT+U functional.  The unusual reaction involves the
simultaneous transfer of $e^-$, breaking of a C-O covalent bond, and the
removal of two oxygen-surface metal coordination.  In 
Fig.~\ref{fig5}e, the C-O bond distance is 1.71~\AA, which is 0.24~\AA\,
shorter than the 1.95~\AA\, predicted using DFT+U.  More conventional 
reactions, such as that C-O bond cleavage reaction in CO$_3^{2-}$, exhibit
$\Delta E^*$ which are less sensitive to choice of hybrid or non-hybrid
DFT functional.\cite{batt}

For CO$_2$ release to occur, the gas molecule must not re-coordinate to
the oxide surface.  In the PBE0 NEB barrier calculation, we accidentally come
across a configuration where the released CO$_2$ adsorbs via another surface
O, reconstituting a CO$_3^{2-}$ motif (Fig.~\ref{fig5}f).  This configuration is
0.49~eV more favorable than Fig.~\ref{fig5}e at zero temperature.  Thus, even
with the gas entropy gained at T=300~K, CO$_2$ release from the (001) surface
of Li$_{0.5}$Ni$_{0.5}$Mn$_{1.5}$O$_{4}$ is arguably barely favorable.

Finally, we examine PBE0 predictions at higher equilibrium voltages.  We lower
the Li-content to 20\% (Li$_{0.2}$Ni$_{0.5}$Mn$_{1.5}$O$_4$) (not shown in
figures), which should raise the equilibrium cathode voltage.  With this
change in Li-content, the blue diamonds in Fig.~\ref{fig4}b indicate that the
last two steps of the reaction are now both exothermic, $\Delta E$=-0.72 and
-0.82~eV, respectively.  The final, CO$_2$-releasing step now exhibits a
much lower $\Delta E^*$=+0.77~eV instead of the +1.22~eV at $x=0.5$.
According to Eq.~\ref{eq1}, CO$_2$ release occurs in sub-second time scales
at $x=0.2$.  Furthermore, the Fig.~\ref{fig5}f-like CO$_2$-adsorption
configuration appropriate to 20\% Li content exhibits a weak CO$_2$
binding energy of $\Delta E$=-0.36~eV, compared to -0.49~eV at $x=0.5$.
In other words, as Mn and Ni cations acquire larger charges, O$^{2-}$ anions
coordinated to them exihbit weaker covalent bonds with CO$_2$ molecules.  This
$\Delta E$ will not retain CO$_2$ on the surface because of the $\sim$0.4~eV
entropy gain upon gas release at T=300~K at 0.1~atm.\cite{ma}

In summary, the more accurate PBE0 method predicts that CO$_2$ release
is not energetically favorable on LMO (001) surfaces at 50\% Li-content.  
Even on LNMO (001) at the same Li-content, which should represent a higher
equilibrium potential than 4.3~V, CO$_2$ release remains energetically and
kinetically hindered.  Only at lower Li-content (20\%, which corresponds
to an even higher equilibrium voltage) does CO$_2$ release become fast.  The
partially oxidized EC fragment is now completely eliminated from the surface
(Fig.~\ref{fig5}c).  Cleared of adsorbed species, the surface can now 
continuously react with solvent molecules (Fig.~\ref{fig1}).  Unlike PBE0,
DFT+U fails to predict this two-step behavior or to distinguish between LMO
and LNMO.

We have not computed the precise CO$_2$ release onset voltage.  This is partly
because the electronic voltage\cite{leung3} is not readily specified in
polaronic conductors like LMO and LNMO (see Sec.~\ref{li2co3} below).  Other
CEI components have been proposed;\cite{jarry} one of them is examined in the
S.I., and found to react at moderate Li-content without releasing CO$_2$.

\subsection{OEMS Measurements}
\label{oems}

\begin{figure}
\centerline{\hbox{ (a) \epsfxsize=3.20in \epsfbox{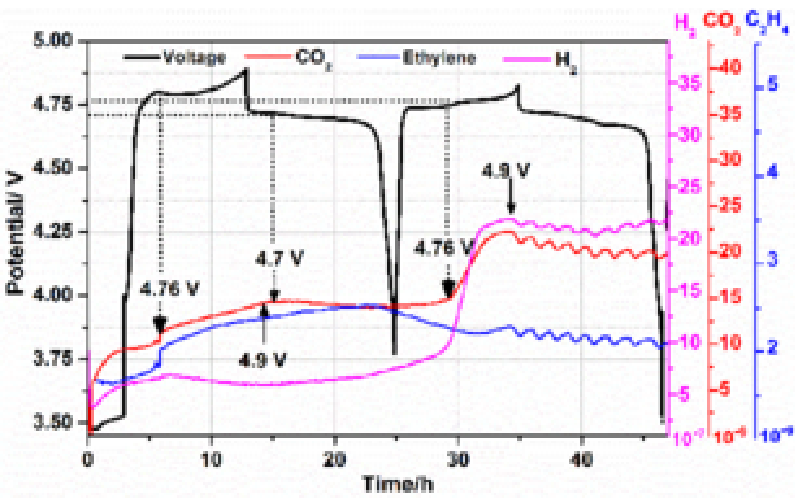} }}
\centerline{\hbox{ (b) \epsfxsize=3.20in \epsfbox{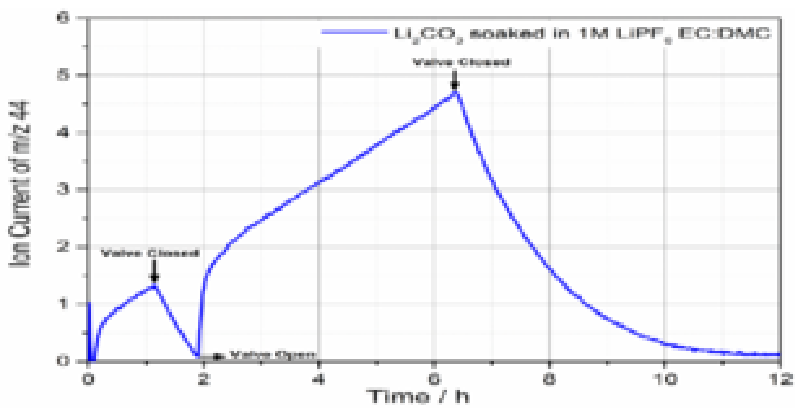} }}
\caption[]
{\label{fig6} \noindent
(a) Galvanostatic charge/discharge profiles (black) and in-operando
online electrochemical mass spectrometric analysis of evolved gases as a
function of applied potential during galvanostatic cycling of a LNMO
half-cell in 1~M LiPF$_6$ in EC:DMC (1:1). (b) OEMS evolution trends
indicating the evolution of CO$_2$ as a byproduct of spontaneous
Li$_2$CO$_3$ dissolution in the electrolyte solution."
}
\end{figure}

To support the theoretical calculations with experimental proofs, we
carried out in-operando analysis of gaseous evolution as a function of
applied potential using online electrochemical mass spectrometry (OEMS)
of LNMO in 1~M LiPF$_6$ in EC:DMC (1:1).  The observed evolution of
H$_2$, CO$_2$, and C$_2$H$_4$ during galvanostatic charge/discharge
is depicted in Fig.~\ref{fig6}a.

The evolution of C$_2$H$_4$ ($m/z$=28) at the beginning of first charge is
likely associated with the SEI formation at the anode side through
reductive decomposition of ethylene carbonate from the electrolytic
solution.\cite{behm}  We assigned the evolution trend at $m/z$=28 to ethylene;
however, partial contribution from CO cannot be eliminated.
By carefully looking at the CO$_2$ evolution, it can be
seen that CO$_2$ starts evolving in the OCV period which can be attributed
to the spontaneous reaction between the electrolyte and the electrode.
This evolution attains equilibrium during the rest period. During
charge to voltages higher than 4.75~V, another evolution of CO$_2$ is
observed. A maximum in the CO$_2$ evolution was observed at 4.9~V. It
is important to note that in OEMS studies, the system is not washed
with inert gas during measurements. This fact enables us to continuously
measure in-operando the volatile components in the battery cell, without
changing its composition and without stopping cascades of reactions
that take place. However, due to this unique approach, remaining gas species
are constantly detected (e.g. CO$_2$); hence our discussion always
focuses on increase/decrease in evolution and local maxima/minima.
During the discharge, when the lithiation level of the LNMO increases
and the potential goes below 4.7~V, the release of CO$_2$ is reduced. 

The onset of CO$_2$ evolution in the second cycle during the
initiation of the second plateau at 4.75~V further confirms the validity
of the calculated results; the evolution of CO$_2$ from CEI decomposition
occurs mainly at lower Li-content and higher voltage.  
Note also that our calculations focus on the EC molecule;
other electrolyte species present in Fig.~\ref{fig6}a, such as DMC, may
yield more gas products during CEI formation at lower voltages. 
H$_2$ also exhibited similar pattern of evolution which indicates close
involvement of H$^+$ in the degradation of electrolyte and evolution of
CO$_2$.  This may also in accordance to the predicted transfer of H$^+$ and
electron in the EC decomposition, with the caveat that the contribution
of anode side to the H$_2$ formation cannot be differentiated.  
We apply OEMS with a half cell (lithium counter/reference electrode). While
OEMS has obvious advantages, our set-up is arguably not well-suited to
pinpoint the source of gases.  At least one recent of modification of
DEMS\cite{behm} seems able to differentiate anode and cathode gas
contributions.  That work also identifies H$_2$, CO, and CO$_2$ products 
from the cathode.

Next we address the question of how LNMO surfaces can become exposed to the
liquid electrolyte despite the fact that as-synthesized cathode oxide materials
tend to be covered with native Li$_2$CO$_3$ films.  We perform experiments
where Li$_2$CO$_3$ particles are soaked in electrolyte with/without LiPF$_6$.
These experiments are
performed under storage conditions, without applied voltages.  We find that
Li$_2$CO$_3$ dissolves in electrolyte solution as well as in pure DMC solvent
(in absence of LiPF$_6$ salt) -- although the latter is much-suppressed.  The
dissolution is qualitatively consistent with previous
studies.\cite{kanno1,aurbach00} In the video included in the S.I., the left
vial is with 0.1~g Li$_2$CO$_3$ in 20~mL of LP30 electrolyte solution, and the
right vial has the same amount of Li$_2$CO$_3$ in 20 mL~DMC (no LiPF$_6$).  In
DMC the Li$_2$CO$_3$ settles down immediately (swirling motion of particles can
be seen), whereas it remains suspended for much longer in the presence
of LiPF$_6$ (LP 30) indicating more gas bubbles adsorbed to Li$_2$CO$_3$
there, and more chance for dissolution.  In neither cases do we see complete
dissolution of the Li$_2$CO$_3$, which is likely in excess.  This finding
suggests that whether native Li$_2$CO$_3$ films on oxide surfaces are
removed upon soaking in the electrolyte depend on the initial film thickness
and electrolyte composition, including impurities present.  Fig.~\ref{fig6}b
depicts OEMS response of Li$_2$CO$_3$ soaked in LP30 solution; it confirms
that CO$_2$ is evolved as a dissolution product.  

\subsection{Li$_2$CO$_3$ Reaction on LNMO surfaces (DFT+U)}
\label{li2co3}

The next two subsections consider oxidation of inorganic Li$_2$CO$_3$
films.\cite{lbl} They suggest alternative mechanisms of Li$_2$CO$_3$ removal
prior to LNMO surfaces reacting with the liquid electrolyte, and complement
our work on organic CEI components.

\begin{figure}
\centerline{\hbox{ (a) \epsfxsize=1.40in \epsfbox{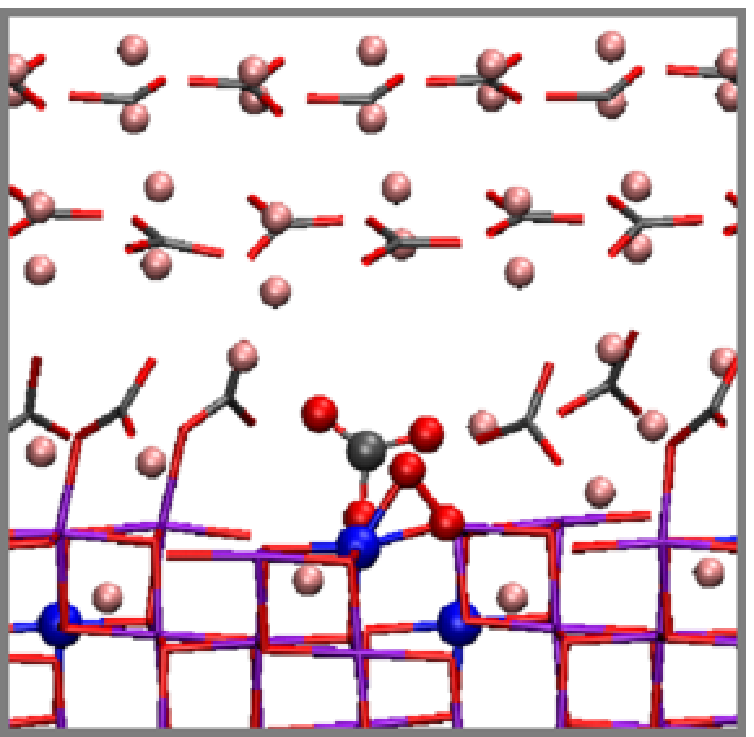} 
                   \epsfxsize=1.450in \epsfbox{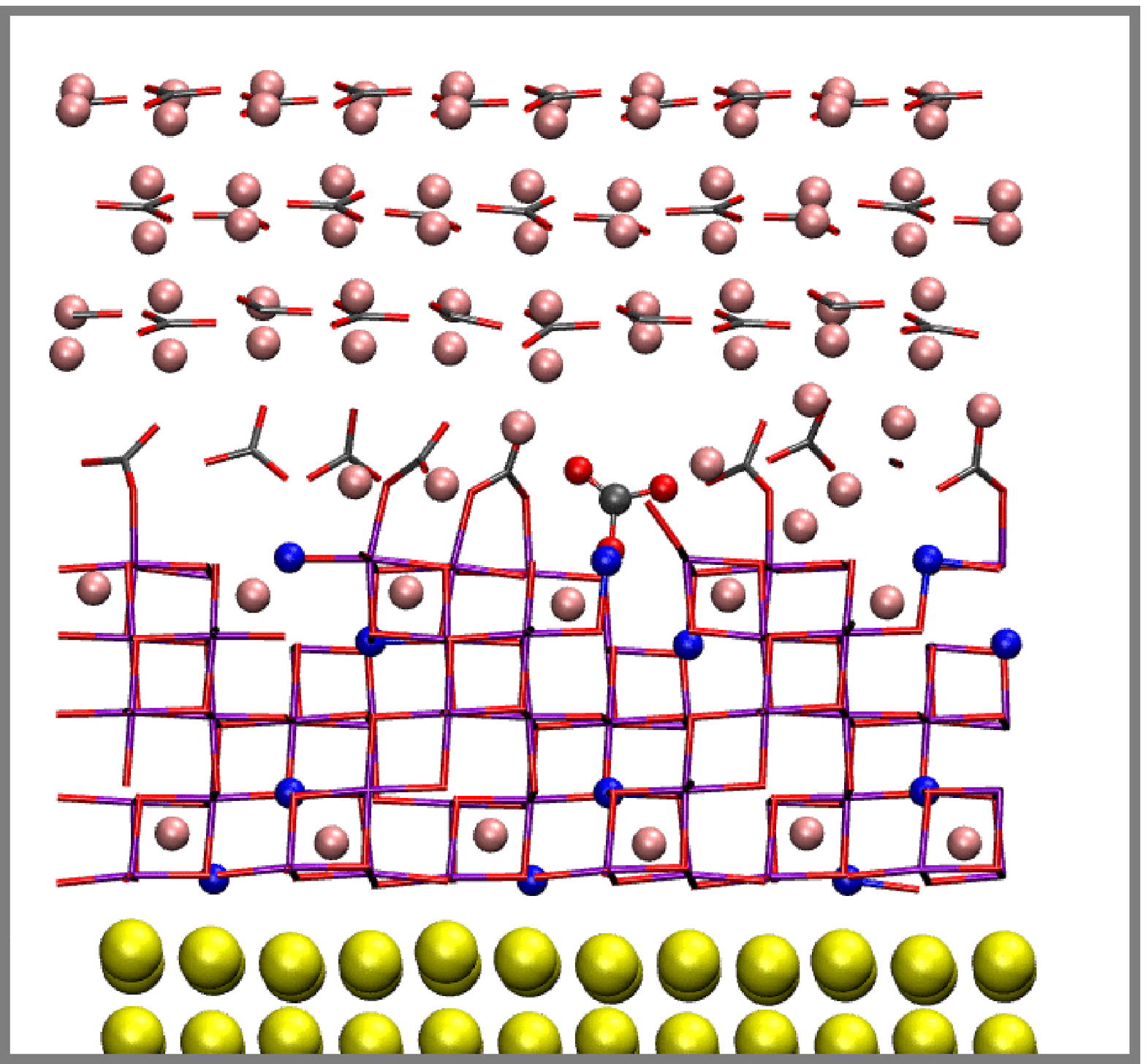} (b) }}
\caption[]
{\label{fig7} \noindent
(a) First reaction products at Li$_2$CO$_3$ (0001)/LNMO (001) interface,
including a O$_2^{\delta-}$; (b) reaction energetics are almost identifical
to panel (a) are predicted when a Au ``current collector'' is added to the
backside of the oxide slab.
}
\end{figure}

In the S.I., we argue that oxidation of Li$_2$CO$_3$ should initiate at
interfaces rather than start from the bulk.  Here we first consider the
interface between its Li$_2$CO$_3$ (0001)\cite{li2co3_1,li2co3_2} and LNMO
(001).  In brief, we find that, to the extent the predicted $\Delta E$ and
$\Delta E^*$ permit oxidation reactions, they are consistent with the release
of triplet O$_2$ gas, not CO$_2$ gas (Fig.~\ref{fig7}a).  This is at variance
with experimental results.\cite{lbl}  Hence we only highlight one
aspect of the results and leave most details to the S.I.

The interface between LNMO (001) and Li$_2$CO$_3$ (Fig.~\ref{fig7}a) is
Li-deficient, and is likely ``metallic.'' Fig.~S3 in the S.I.\, confirms the
interfacial region has no band gap.  Due to this metallic behavior, the
electronic voltage ($V_e$) can be readily computed as the LNMO work function
modified by its interface with Li$_2$CO$_3$ via the Trasatti
relation,\cite{trasatti86,leung3,note2} provided that the electrolyte 
contribution is neglected. The computed
$V_e$ and the voltage based on Li-insertion energetics must match to ensure
that simulation cells containing interfaces are at ``equilibrium,'' not at
overpotential conditions.  We obtain an electronic voltage of $V_e$=4.56~V.
This value is reasonably close to the 4.75~V found by adding/subtracting
Li-atoms.  In other words, the model system is not at a significant
overpotential.  A more general way to compute $V_e$, even when the electrode
has a band gap, is to place an inert metallic electrode underneath.\cite{leung3}
Upon adding a Au (001) ``current collector'' (Fig.~\ref{fig7}b), $V_e$ is
found to an almost indistinguishable 4.58~V.  Therefore the metallic nature
of the thin interface between LNMO and Li$_2$CO$_3$ should be sufficent for
establishing the electronic potential.  The $\Delta E$ of the reaction
associated with CO$_3^{2-}$ reaction to form
O$_2^{\delta-}$ is also very similar to the slab model without the Au slab.
This dovetails with our experience that redox reactions on cathode surfaces,
which involve discrete changes in occupancies of transition metal ion
$d$-orbitals at orbital energies below the Fermi level (S.I.), are not
sensitive to $V_e$,\cite{leung3} unlike reactions on graphitic
anodes.\cite{leung4}  Therefore we have not made further
modifications of the interfaces to bring $V_e$ closer to 4.75~V.  

There is no metallic character to the LNMO slabs in the previous sections.
The cost of PBE0 calculations there makes adding a metallic
``current collector'' beneath the LNMO (001) difficult, and this impedes
effort to calculate $V_e$.  Along with the small cell sizes, this is one
reason we have not reported CO$_2$ release onset voltages.

\subsection{Reaction between Li$_2$CO$_3$ and Solvent Molecules}
\label{li2co3_ec}

\begin{figure}
%\centerline{\hbox{ (a) \epsfxsize=1.50in \epsfbox{fig8a.ps} 
%                   (b) \epsfxsize=1.50in \epsfbox{fig8b.ps} }}
%\centerline{\hbox{ (c) \epsfxsize=1.50in \epsfbox{fig8c.ps} 
%                   (d) \epsfxsize=1.50in \epsfbox{fig8d.ps} }}
%\centerline{\hbox{ (e) \epsfxsize=1.50in \epsfbox{fig8e.ps} }}
\centerline{\hbox{  \epsfxsize=3.00in \epsfbox{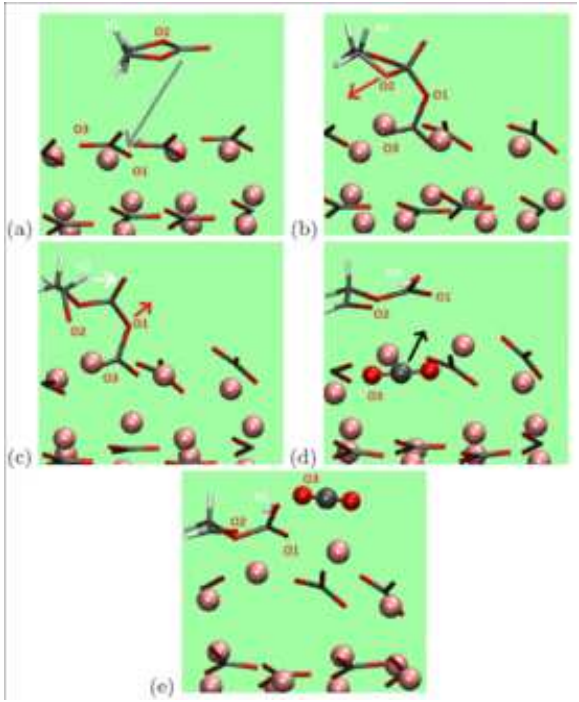} }}
\caption[]
{\label{fig8} \noindent
EC oxidative decomposition on a partially delithated Li$_2$CO$_3$ (0001)
surface.  (a) Physisorbed configuration; (b) EC adsorbed in a bent
geometry; (c) broken EC C-O bond; (d) proton transfer from EC to 
lithium carbonate and breaking of O-C bond, forming a CO$_2$;
(e) release of CO$_2$ as gas molecule.
}
\end{figure}

Finally, we explore electrolyte oxidative reactions on the outer, partially
delithiated Li$_2$CO$_3$ (0001) surface.  The PBE functional is applied.  As
no transition metal ion is present in this section, DFT+U and PBE are
equivalent. 

Fig.~\ref{fig8}a depicts the (0001) surface with a single physisorbed EC
molecule.  There should be 8~Li$^+$ on each layer of Li$_2$CO$_3$ in the
simulation cell periodically replicated in the lateral dimensions.  The top
layer has half its Li atoms removed to qualitatively mimic high voltage
conditions.  The simulation cell remains charge-neutral.

\begin{figure}
\centerline{\hbox{ \epsfxsize=3.00in \epsfbox{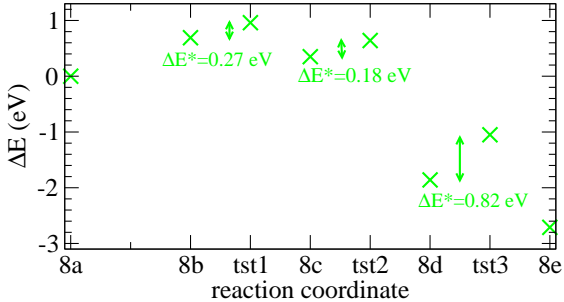} }}
\caption[]
{\label{fig9} \noindent
PBE relative energies and energy barriers, in eV, associated with EC oxidation
on partially delithiated Li$_2$CO$_3$ (0001) surfaces (Fig.~\ref{fig8}).
The $x$-axis labels refer to the panels in Fig.~\ref{fig8}.  The three
transition states are not depicted in Fig.~\ref{fig8}.
}
\end{figure}

We explore a reaction mechanism similar to that for EC oxidation on LMO (001)
(Fig.~\ref{fig2}).\cite{leung1}  In Fig.~\ref{fig8}b, the EC is chemisorbed
with a bent molecular geometry.  This is endothermic by $\Delta E$=+0.69~eV
(Fig~\ref{fig9}).  Such a configuration change should exhibit no
barrier.\cite{leung1} The carbonyl carbon atom is now 4-coordinated.
Fig.~\ref{fig8}c depicts a subsequent configuration with a broken C-O bond,
opening the EC 5-member ring and restoring 3-coordination to what was the
carbonyl C-atom.  The reaction is exothermic by -0.33~eV compared with the
bent configuration (Fig.~\ref{fig8}b), but endothermic by +0.35~eV relative to
physisorption (Fig.~\ref{fig8}a).  The barrier is $\Delta E^*$=0.+27~eV
relative to Fig.~\ref{fig8}b (+0.96~eV relative to Fig.~\ref{fig8}a).

The next step is postulated to involve the transfer of a ethylene proton to
a nearby oxygen group, in accordance with EC oxidation on oxide
surfaces.\cite{leung1} Unlike LMO or LNMO (001), O~atoms on the Li$_2$CO$_3$
(0001) surface are far away from all protons.  When an H-atom is moved from
an EC fragment C atom to a surface CO$_3^{2-}$ group and the configuration is
optimized, we find that the proton spontaneous migrates to an oxygen atom
on the ROCO$_2^{2-}$ group on the EC fragment instead (Fig.~\ref{fig8}d).
Simultaneously, a C-O bond is broken, yielding a CO$_2$ on the Li$_2$CO$_3$
surface.  This H-transfer step is accompanied by $e^-$ transfer (oxidation)
and is overall oxothermic, $\Delta E$=-1.86~eV, relative to physisorption.
The barrier is $\Delta E$=0.18~eV relative to Fig.~\ref{fig8}c.
In Fig.~\ref{fig8}e, the CO$_2$ lodged on the surface is released alongside
an additional $\Delta E$=-0.84~eV with a local $\Delta E^*$=+0.82~eV barrier.
Overall, the highest activation energy among these steps is 0.96~eV.  These
reactions at the electrolyte/Li$_2$CO$_3$ interface yield more favorable
$\Delta E$ and $\Delta E^*$ than at the LNMO/Li$_2$CO$_3$ interface (S.I.).

This reaction pathway represents an alternative to that suggested in
Ref.~\onlinecite{mahne2018}, namely that singlet O$_2$ is released during
oxidationwhich then attacks the liquid electrolyte.
We stress that this section represents an exploratory plausibility
demonstration.  We have not exhaustively explored other mechanisms,
nor have we applied the PBE0 functional to check $\Delta E^*$.  The precise
voltage/lithium content relation for this reaction has not been determined.
Despite these caveats, this section suggests that EC reactions with
surface-delithiated Li$_2$CO$_3$ to release CO$_2$ are viable.
Even accounting for possible DFT+U $\Delta E^*$
errors, sufficiently high voltage will likely make the EC/Li$_x$CO$_3$
reaction sufficiently favorable to occur in one-hour time scales, just
like it does for EC fragment oxidation on LNMO surfaces.  

%\begin{table}
%\begin{tabular}{||l|l|l|l|l|l||} \hline
%config. & Fig.~\ref{fig8}a & Fig.~\ref{fig8}b & Fig.~\ref{fig8}c &
%		Fig.~\ref{fig8}d & Fig.~\ref{fig8}e  \\ \hline
%$\Delta E$ (eV) & 0.00 & 0.69 & -0.33  & -2.24  & -0.84 \\
%                & 0.00 & (0.69) &  (0.35) & (-1.86) & (-2.71) \\ \hline
%$\Delta E^*$ (eV) & &   & 0.96 &  0.18   & +0.82   \\
%	      &  &  & (0.96) &  (0.64)   & (-1.05) \\
%\hline
%\end{tabular}
%\caption[]
%{\label{table2} \noindent
%Relative DFT/PBE energies ($\Delta E$) and energy barriers ($\Delta E^*$),
%in eV, associated with EC oxidation on partially delithiated Li$_2$CO$_3$
%(0001) surfaces (Fig.~\ref{fig8}).  $\Delta E^*$ are listed under the product
%configuration; the reactant is directly to its left.  Each unbracketted value
%is the difference between the configuration and the intermediate immediately
%preceding it; the bracketted one is between product and the Fig.~\ref{fig8}a
%physisorbed configuration.  
%}
%\end{table}

\subsection{Further Discussions}
\label{further}

The further oxidation of our model CEI products involves the simultaneous
transfer of H$^+$ and $e^-$.  The role of proton transfer in enabling oxidation
of organic components has been emphasized in the presence of counter ions and at
multiple battery interfaces;\cite{borodin1,leung1,leung2,borodin2015,giordano1,giordano2,musgrave1,tebbe,morgan,bal19,li17} 
its importance cannot be overstated.  One product of our predicted oxidative
decomposition is C$_2$H$_2$O$_2$, known to be a polymerizing agent.  This
appears consistent with polymeric CEI species routinely reported on cathode
surfaces,\cite{edstrom,aurbach01,aurbach99} although other mechanisms can 
lead to polymers.

Oxidation of organic molecules and fragments on cathode oxide surfaces
inject electrons into transition metal ions therein.  To offset this
electrochemical reduction and to maintain a constant
potential, Li atoms should be removed (with Li$^+$ diffusing into the
electrolyte).  Ideally, the grand canonical Monte Carlo (GCMC) technique
should be combined with DFT calculations for this purpose; so far DFT-based
MC is in its infancy.\cite{sugino}

We propose a multi-step electrolyte oxidation pathway (Fig.~\ref{fig1}a),
which is also suggested for oxidation on Pt surfaces.\cite{cei_pt}
Electrolytes with fluorinated solvent molecules
exhibits increased anodic stability.\cite{borodin_nature,shkrob2017}
This behavior is not inconsistent with our hypothesis, because CEI formed
from fluorinated molecules should also be more stable against
oxidation than CEI formed from standard organic battery electrolytes.  This
hypothesis is not inconsistent with 
Ref.~\onlinecite{tateyama}, which predicts a one-step release of CO$_2$ from
intact EC molecules adsorbed on a pristine LNMO (001) surface.  Such a direct
CO$_2$ release route could take place in a voltage regime where all organic CEI
products have been oxidized and removed from LNMO surfaces.

The reported voltage onset associated with CO$_2$ release varies in the
literature.  He {\it et al.},\cite{novak2016} Michalak
{\it et al.},\cite{janek2016} and Xu {\it et al.}\cite{gasteiger2014} have
reported CO$_2$ release below 5.0~V (4.8~V, 4.6~V, and 4.8~V respectively),
while Jusys {\it et al.}\cite{behm} and Jung {\it et al.}\cite{gasteiger2017}
have reported $\sim$5.0~V and $>$5.4~V, respectively.  Our reported OEMS
results are meant to be a check on this issue.  We report 4.75~V
(Fig.~\ref{fig6}), which is in alignment with
Refs.~\onlinecite{novak2016,janek2016,gasteiger2014}.  The differences
may arise from charging rates, anode choices,\cite{abraham} electrolytes
impurities,\cite{metzger2016a} temperatures, and the initial surface
termination of cathode oxide materials.  Interestingly, high-Ni content
layered NMC materials are reported to release CO$_2$ at much lower voltages
than LNMO.\cite{gasteiger2017}  In another report,\cite{novak2016}
DMC is found to be more reactive than EC on LNMO.  Our calculations have
focused on LNMO (001) due to the small cell size that permits PBE0-based
calculations, and on EC because we have performed previous
studies.\cite{leung1,leung2}  In the future, extending our method to DMC,
NMC, counter-ions, ond the (111) facet will give interesting comparisons.

CO release hve been reported at elevated voltages.  
CO cannot emerge from the mechanism investigated herein, and will
be considered in the future.

\section{Conclusions}
\label{conclusions}

We have applied the hybrid PBE0 DFT functional to compute the energetics
and kinetics associated with key steps of interfacial oxidation of model
cathode electrolyte interphase (CEI) components on high voltage spinel
(Li$_{x}$Ni$_{0.5}$Mn$_{1.5}$O$_4$ or LNMO) (001) surfaces.  Our model CEI
components are partially oxidized ethylene carbonate (EC) molecules from
previous computational studies.\cite{leung1}  At moderate Li-content ($x$=0.5),
the oxidative reaction barrier $\Delta E^*$ is too high.  At much lower
Li-content ($x$=0.2) which corresponds to higher equilibrium voltages,
$\Delta E^*$ decreases, and the oxidative reaction occurs within 1-hour
battery charging time scales.  This leads to removal of adsorbed organic
fragments from this surface and release of CO$_2$ molecules.
The precise onset voltage cannot yet be determined.  Spinel oxides not doped
with Ni, i.e., Li$_{x}$Mn$_2$O$_4$ (LMO), exhibits reaction energies
($\Delta E$) far less favorable towards CO$_2$ release compared with LNMO.

We also apply the DFT/PBE0 method to re-examine oxidation kinetics of intact
EC molecules on LMO (001) surfaces at 40\% charge ($x$=0.6).  $\Delta E$ and
$\Delta E^*$ for the key oxidative step are predicted to be -1.75~eV
and 1.05~eV, respectively, to yield adsorbed species which are the 
CEI components discussed in the last paragraph.  This 1.05~eV barrier is much
lower than the $\Delta E^*$ for a subsequent CEI oxidation reaction at similar
Li-content ($x$=0.5), and is further reduced by zero point corrections.

From these predictions, our PBE0 calculations are consistent with
a two-step process.  First EC solvent molecules are oxidized at modest voltages
and Li-contents.  The partially decomposed EC fragment remains on the LMO and
LNMO (001) surfaces, covering up the reactive transition metal ion sites.
At sufficiently high voltages on LNMO (001), these fragments are oxidized,
releasing CO$_2$ gas and clearing the surface for further, uncontrolled
reactions with the liquid electrolyte.  LMO likely behaves similarly at 
sufficiently high voltages, but we have not demonstrated this expliclty.
Fluoride- and phosphorus-containing CEI products have not been considered in
this work.

The widely applied DFT+U method, based on the PBE functional, is useful for 
predicting qualitative oxidative mechanisms.  However, unlike PBE0, it predicts
$\Delta E$ and $\Delta E^*$ which are favorable for oxidation of both EC and
adsorbed, partially decomposed EC fragments -- even at the modest equilibrium
potentials associated with Li$_{0.5}$Mn$_2$O$_4$.  This does not appear to
agree our online electrochemical mass spectroscopy (OEMS) measurements and
those of other groups.  This strongly suggests that hybrid DFT functionals
should be used to spot-check electrolyte oxidation predictions.  Nevertheless,
we have used the more economic DFT+U method for an exploratory investigation
of oxidation of native Li$_2$CO$_3$ films on cathode oxides.  We conclude that
oxidation of Li$_2$CO$_3$ is more likely to first occur on the outer
Li$_2$CO$_3$ surface in contact with liquid electrolytes, than on its inner
surface in contact with cathode oxide materials.

All DFT/PBE0 calculations of the model slabs without Li$_2$CO$_3$ films
yield a band gap.  Hence there
are no delocalized ``surface states'' -- the redox-active states are simply
$d$-orbitals localized on surface Mn and Ni transition metal ions.  Higher
voltages yield more Ni(IV), Ni(III), and Mn(IV) cations and accelerate organic
CEI degradation reactions.  Two general conclusions that can be drawn from
our specific PBE0 functional calculations are: PBE0 is probably more
accurate than DFT+U when applied to Mn(II)/Mn(III) redox couple. PBE0
predicts higher reaction barriers associated with C-H and C-O bond-breaking,
especially in reactions that releases CO$_2$. 

This comparative study on EC, organic CEI component, and Li$_2$CO$_3$ oxidation
highlights the importance of multi-step reactions, and emphasizes the need to
examine the oxidation of CEI/surface films, not just intact solvent molecules.
Differentiating CEI and solvent oxidation events under high voltage conditions
should lead to new insights that inform cathode passivation strategies.

\section*{Supplementary Material}

See supplementary material for simulation cell size effects; the
reactions of Li$_2$CO$_3$ on LNMO (001); reactions between EC molecules
and other oxide surfaces; oxidation of other proposed CEI components,
rationale for using PBE0; rationale for single molecule slab models;
optimized PBE0 configurations; and a video showing the reaction
between Li$_2$CO$_3$ and liquid electrolytes under storage conditions.
 
\section*{Acknowledgements}
 
We thank Shen Dillon for careful reading of an early draft, and Angelique
Jarry, Dale Huber, Jacob Harvey, Katharine Harrison, Christine James, and
Yue Qi for useful discussions.  MN would like to acknowledge the Funding of
Israel science foundation (Grant no.~2028/17 and 2209/17) and support of
Planning Budgeting Committee/ISRAEL Council for Higher Education (CHE) and
Fuel Choice Initiative (Prime Minister Office of ISRAEL), within
the framework of Israel National Research Center for Electrochemical
Propulsion (INREP).  KL was supported by Nanostructures for Electrical
Energy Storage (NEES), an Energy Frontier Research Center funded by
the U.S. Department of Energy, Office of Science, Office of Basic
Energy Sciences under Award Number DESC0001160.  Sandia National
Laboratories is a multimission laboratory managed and operated by
National Technology and Engineering Solutions of Sandia, LLC, a
wholly owned subsidiary of Honeywell International, Inc., for the
U.S. Department of Energy's National Nuclear Security Administration
under contract DE-NA0003525.  This paper describes objective technical
results and analysis. Any subjective views or opinions that might be
expressed in the paper do not necessarily represent the views of the
U.S. Department of Energy or the United States Government.

%\include{ref}

%\newpage

%\begin{figure*}
%\centerline{\hbox{ \epsfxsize=6.00in \epsfbox{toc.ps} }}
%\caption[]
%{\label{toc} \noindent
%Table of content figure.
%}
%\end{figure*}


\begin{references}
\bibitem{note1}
Such CEI formation behavior would be distinct from lithium ion battery
anodes where parasitic reactions responsible for formation and evolution
of solid-electrolyte-interphase surface films often involve long-range
electron transfer, and can occur away from electrode surfaces.
\bibitem{lmo111}
The (111) surface should be most prominent, and reconstructed (111)
surface models have been devised.\cite{persson14,persson13,vitaly}  
Direct contact betweeh molecules and transition metal ions on the (111)
surface will first require creation of oxygen vacancies.  The (001)
surface, which requires a smaller simulation cell and is commensurate with
Li$_2$CO$_3$ lattice constants, is adopted herein for ease of computation.
\bibitem{note2}
The potential should technically also depends on the outer interface between
Li$_2$CO$_3$ and the polymeric CEI on top of it, as well as the polymeric
CEI/liquid electrolyte interface.  The structures of these highly
complex interfaces are unknown, and they are neglected herein.
\end{references}
\end{document}